\documentclass[twocolumn,showpacs,prl,amsfonts,amsmath,amssymb,floatfix]{revtex4} 
\usepackage{color}
\usepackage{hhline}
\usepackage{mathrsfs}
\usepackage{graphicx}
\usepackage{dcolumn}
\usepackage{bm}
\usepackage{multirow}
\usepackage{booktabs}
\usepackage{afterpage}
\usepackage{amsmath}

\begin{document}

\title{Development of Density-Functional Theory for a Plasmon-Assisted Superconducting State: Application to Lithium Under High Pressures}

\author{Ryosuke Akashi$^{1}$}
\author{Ryotaro Arita$^{1,2}$}
\affiliation{$^1$Department of Applied Physics, The University of Tokyo, 
             Hongo, Bunkyo-ku, Tokyo 113-8656, Japan}
\affiliation{$^2$JST-PRESTO, Kawaguchi, Saitama 332-0012, Japan}

\date{\today}
\begin{abstract}
We extend the density-functional theory for superconductors (SCDFT) to take account of the dynamical structure of the screened Coulomb interaction. We construct an exchange-correlation kernel in the SCDFT gap equation on the basis of the random-phase approximation, where electronic collective excitations such as plasmons are properly treated. Through an application to fcc lithium under high pressures, we demonstrate that our new kernel gives higher transition temperatures ($T_{\rm c}$) when the plasmon and phonon cooperatively mediate pairing and it improves the agreement between the calculated and experimentally observed $T_{\rm c}$. The present formalism opens the door to non-empirical studies on unconventional electron mechanisms of superconductivity based on density functional theory.
\end{abstract}

\maketitle

{\it -Introduction.} The electron mechanism of superconductivity has long been a central topic of theoretical physics. Since the seminal discovery of cuprate superconductors, correlated systems with short-range Coulomb interaction have been extensively studied. On the other hand, systems with long-range Coulomb interaction has also a long history, dating back to Kohn and Luttinger~\cite{Kohn-Luttinger}. Particularly, there have been several theoretical proposals of superconducting mechanisms that exploit dynamical structure of the screened Coulomb interaction, which is represented by the frequency dependence of the dielectric function, $\varepsilon(\omega)$. In fact, even without the electron-phonon interactions, it has been shown that superconductivity is induced by the dynamical Coulomb interactions due to exchange of plasmons~\cite{Takada1978,Rietschel-Sham1983} or excitons~\cite{Little1967}. More interestingly, the dynamical electron-electron interaction can cooperate with the ordinary electron-phonon pairing interaction. This possibility has been investigated for SrTiO$_{3}$~\cite{Koonce-Cohen1967, Takada-SrTiO3}, $s$-$d$ transition metals~\cite{Garland-sd}, insulator-sandwiched systems~\cite{Ginzburg-HTSC,ABB1973}, and layered systems where soft electronic collective modes exist~\cite{Kresin1987}. Moreover, recent discoveries of high-temperature superconductivity in doped band insulators have indeed stimulated further studies on this issue~\cite{Yamanaka1998,Bill-rapid2002,Taguchi2006,Ye2012}.

One important aspect of the above issues is that one must deal with quantitative competition between the attractive and repulsive contributions of the screened Coulomb interaction. Here, first-principles studies based on density functional theory are expected to provide deep insight. Recently, the progress in density functional theory for superconductors (SCDFT)~\cite{GrossI,GrossII} has enabled us to calculate superconducting transition temperature ($T_{\rm c}$) nonempirically. Although the current SCDFT treats the screened electron-electron interaction within the static level, it achieves extreme accuracy for various conventional superconductors~\cite{GrossII,Floris-MgB2,Sanna-CaC6} by including the dynamical phonon effect referring to the Migdal-Eliashberg (ME) theory~\cite{Migdal-Eliashberg}. However, to formulate a quantitative \textit{ab initio} method for unconventional plasmon or exciton mechanism, we need to go beyond the current SCDFT with the static approximation for the screened electron-electron interaction.

In this paper, we extend the current SCDFT and develop a scheme to calculate $T_{\rm c}$ with considering the dynamical structure of the screened Coulomb interaction. Our formulation provides a general treatment of collective excitations described by the zeros of $\varepsilon(\omega)$, that is, plasmons in solids. Our scheme enables stable calculations considering the frequency dependence of the screened Coulomb interaction, and is applicable to superconductors regardless of whether the plasmon effect is significant or not. As a representative system, we study superconducting fcc lithium under high pressure~\cite{Shimizu2002,Struzhkin2002,Deemyad2003}, which is expected to be similar to the electron gas with a small amount of carriers studied in Refs.~\onlinecite{Takada1978} and \onlinecite{Rietschel-Sham1983}. The previous calculation employing the current SCDFT~\cite{Profeta-pressure} reproduced the experimental $T_{\rm c}$ with the calculated electron-phonon coupling $\lambda$$>$$1$ under the pressure $>$$20$GPa, whereas the most recent \textit{ab initio} phonon calculation~\cite{Bazirov-pressure} gave far smaller $\lambda$$\lesssim$$0.8$. We show that, even with $\lambda$$\lesssim$0.8, the $T_{\rm c}$ calculated with the plasmon contribution agrees well with the experimentally observed $T_{\rm c}$.

\begin{figure}[t]
 \begin{center}
  \includegraphics[scale=.2]{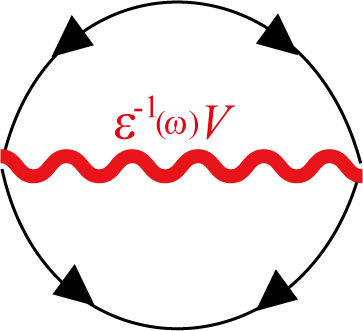}
  \caption{(Color online) Diagram corresponding to the electron nondiagonal kernel, $\mathcal{K}^{\rm el}$. The solid line with arrows running in the opposite direction denotes the electronic anomalous propagator~\cite{GrossI}. The red wavy line denotes the screened electronic Coulomb interaction, which is a product of the inverse dielectric function $\varepsilon^{-1}$ and the bare Coulomb interaction $V$.}
  \label{fig:diagram}
 \end{center}
\end{figure}

{\it -Formulation.} In the current SCDFT~\cite{GrossI, GrossII}, $T_{\rm c}$ is obtained by solving the gap equation
\begin{eqnarray}
\Delta_{n{\bf k}}\!=\!-\mathcal{Z}_{n\!{\bf k}}\!\Delta_{n\!{\bf k}}
\!-\!\frac{1}{2}\!\sum_{n'\!{\bf k'}}\!\mathcal{K}_{n\!{\bf k}\!n'{\bf k}'}
\!\frac{\mathrm{tanh}[(\!\beta/2\!)\!E_{n'{\bf k'}}\!]}{E_{n'{\bf k'}}}\!\Delta_{n'\!{\bf k'}}.
\label{eq:gap-eq}
\end{eqnarray}
Here, $n$ and ${\bf k}$ denote the band index and crystal momentum, respectively, $\Delta$ is the gap function, and $\beta$ is the inverse temperature. The energy $E_{n {\bf k}}$ is defined as $E_{n {\bf k}}$=$\sqrt{\xi_{n {\bf k}}^{2}+\Delta_{n {\bf k}}^{2}}$ and $\xi_{n {\bf k}}=\epsilon_{n {\bf k}}-\mu$ is the one-electron energy measured from the chemical potential $\mu$, where $\epsilon_{n {\bf k}}$ is obtained by solving the normal Kohn-Sham equation in density functional theory
$
\mathcal{H}_{\rm KS}|\varphi_{n{\bf k}}\rangle=\epsilon_{n{\bf k}}
|\varphi_{n{\bf k}}\rangle
$
with $\mathcal{H}_{\rm KS}$ and $|\varphi_{n{\bf k}}\rangle$ being the Kohn-Sham Hamiltonian and the Bloch state, respectively. The functions $\mathcal{Z}$ and $\mathcal{K}$ are the exchange-correlation kernels describing the effects of the interactions. The nondiagonal kernel $\mathcal{K}$ is composed of two parts $\mathcal{K}$$=$$\mathcal{K}^{\rm ph}$$+$$\mathcal{K}^{\rm el}$ representing the electron-phonon and electron-electron interactions, whereas the diagonal kernel $\mathcal{Z}$ consists of one contribution $\mathcal{Z}$$=$$\mathcal{Z}^{\rm ph}$ representing the mass renormalization of the normal-state bandstructure due to the electron-phonon coupling. The phonon parts, $\mathcal{K}^{\rm ph}$ and $\mathcal{Z}^{\rm ph}$, have been formulated so that the conventional strong-coupling superconductivity can be properly treated. The electronic nondiagonal kernel $\mathcal{K}^{\rm el}$ corresponds to the diagram depicted in Fig.~\ref{fig:diagram} with the static approximation $\varepsilon(\omega)$$\simeq$$\varepsilon(0)$. We go beyond this approximation by retaining its frequency dependence. The diagram thus yields a new form for $\mathcal{K}^{\rm el}$ as
\begin{eqnarray}
&&
\mathcal{K}^{\rm el, dyn}_{n{\bf k},n'{\bf k}}
\!=\!
\lim_{\{\Delta_{n{\bf k}}\}\rightarrow 0}
\frac{1}{{\rm tanh}[(\beta /2 ) E_{n{\bf k}}]}
\frac{1}{{\rm tanh}[(\beta /2) E_{n'{\bf k}'}]}
\frac{1}{\beta^{2}}
\nonumber \\
&&
\hspace{10pt}\times
\sum_{\omega_{1}\omega_{2}}
F_{n{\bf k}}({\rm i}\omega_{1})
F_{n'{\bf k}'}({\rm i}\omega_{2})
W_{n{\bf k}n'{\bf k}'}[{\rm i}(\omega_{1}\!\!-\!\!\omega_{2})]
,
\label{eq:kernel-dyn}
\end{eqnarray}
where $F_{n{\bf k}}({\rm i}\omega)$
$=$
$\frac{1}{{\rm i}\omega\!+\!E_{n{\bf k}}}
\!-\!
\frac{1}{{\rm i}\omega\!-\!E_{n{\bf k}}}
$
 and $\omega_{1}$ and $\omega_{2}$ denote the fermionic Matsubara frequency. Function $W_{n{\bf k}n'{\bf k}'}({\rm i}\omega)$$\equiv$$\langle \varphi_{n{\bf k}\uparrow}\varphi_{n-{\bf k}\downarrow}|\varepsilon^{-1}({\rm i}\omega)V|\varphi_{n'{\bf k}'\uparrow}\varphi_{n'-{\bf k}'\downarrow}\rangle$ denotes the screened Coulomb interaction scattering the Cooper pairs.

In order to treat the plasmon-induced $\omega$ dependence of the screened Coulomb interaction, we employ the random-phase approximation (RPA~\cite{RPA}) for $\varepsilon(\omega)$. Further, we analytically carry out the Matsubara summation in Eq.~(\ref{eq:kernel-dyn}) by approximating $W_{n{\bf k}n'{\bf k}'}({\rm i}\omega)$ as a simple analytic function $\tilde{W}_{n{\bf k}n'{\bf k}'}({\rm i}\omega)$. We employ a multi-pole plasmon approximation represented as
\begin{eqnarray}
&&
\tilde{W}_{n{\bf k}n'{\bf k}'}({\rm i}\nu_{m})
\!=\!
W_{n{\bf k}n'{\bf k}'}(0)
\nonumber \\
&&\hspace{10pt}
\!+\!
\sum^{N_{\rm p}}_{i}
a_{i;n{\bf k}n'{\bf k}'}
\left[
\frac{2}{\omega_{i;n{\bf k}n'{\bf k}'}}
-\frac{2\omega_{i;n{\bf k}n'{\bf k}'}}{\nu_{m}^{2}\!+\!\omega^{2}_{i;n{\bf k}n'{\bf k}'}}
\right]
,
\label{eq:plasmon-pole}
\end{eqnarray}
where $N_{\rm p}$ is the number of plasmon modes and $\nu_{m}$ denotes the bosonic Matsubara frequency. The functions $a_{i;n{\bf k}n'{\bf k}'}$ and $\omega_{i;n{\bf k}n'{\bf k}'}$ are the plasmon coupling and the plasmon frequency of the $i$-th mode, respectively. We have formulated so that $\tilde{W}_{n{\bf k}n'{\bf k}'}(0)$$=$$W_{n{\bf k}n'{\bf k}'}(0)$. Substituting Eq.~(\ref{eq:plasmon-pole}) in Eq.~(\ref{eq:kernel-dyn}), we finally obtain $\mathcal{K}^{\rm el,dyn}$$=$$\mathcal{K}^{\rm el,stat}$$+$$\Delta\mathcal{K}^{\rm el}$ with $\mathcal{K}^{\rm el,stat}_{n{\bf k}n'{\bf k}'}$$=$$W_{n{\bf k}n'{\bf k}'}(0)$ and
\begin{eqnarray}
\Delta\mathcal{K}^{\rm el}_{n{\bf k},n'{\bf k}}
&=&
\sum_{i}^{N_{\rm p}}\!2a_{i;n{\bf k}n'{\bf k}'} \!\left[
\frac{1}
{\omega_{i;n{\bf k}n'{\bf k}'}} \right.
\nonumber \\
&&
\hspace{-50pt}
\left.
+
\frac{
I\!(\xi_{n{\bf k}}\!,\!\xi_{n'{\bf k}'}\!,\omega_{i;n{\bf k}n'{\bf k}'}\!)
\!\!-\!\!
I\!(\xi_{n{\bf k}}\!,-\!\xi_{n'{\bf k}'}\!,\omega_{i;n{\bf k}n'{\bf k}'}\!)
}{{\rm tanh}[(\beta/2) \xi_{n{\bf k}}]{\rm tanh}[(\beta/2) \xi_{n'{\bf k}'}]}
\right]
,
\label{eq:Delta-kernel}
\end{eqnarray}
where the function $I$ is defined by Eq.~(55) in Ref.~\onlinecite{GrossI}.

\begin{figure}[b]
 \begin{center}
  \includegraphics[scale=.65]{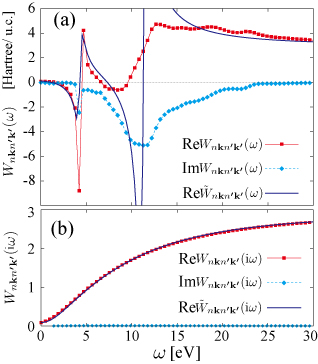}
  \caption{(Color online) Screened Coulomb interaction $W_{n{\bf k}n'{\bf k}'}$ and the corresponding approximate function $\tilde{W}_{n{\bf k}n'{\bf k}'}$ for fcc lithium in 14GPa calculated along the real frequency axis [(a)], and the imaginary frequency axis [(b)]. The band indices $n$ and $n'$ correspond to the partially occupied band, ${\bf k}$$=$$(2\pi/a)(1/7,1/7,1/7)$, and ${\bf k}'$$=$$(0,0,0)$.}
  \label{fig:fit}
 \end{center}
\end{figure}

We determine $(a_{i;n{\bf k}n'{\bf k}'}, \omega_{i;n{\bf k}n'{\bf k}'})$ in the following procedure: (i) calculate the screened Coulomb interaction for the {\it real} frequency grid $W_{n{\bf k}n'{\bf k}'}(\omega\!+\!{\rm i}0^{+})$, (ii) determine the plasmon frequencies $\{\omega_{i;n{\bf k}n'{\bf k}'}\}$ by the position of the peaks up to the $N_{\rm p}$-th largest in ${\rm Im}W_{n{\bf k}n'{\bf k}'}(\omega\!+\!{\rm i}0^{+})$, (ii) calculate the screened Coulomb interaction for the {\it imaginary} frequency grid $W_{n{\bf k}n'{\bf k}'}({\rm i}\omega)$, and (iv) using the calculated $W_{n{\bf k}n'{\bf k}'}({\rm i}\omega)$, determine the plasmon coupling coefficients $\{a_{i;n{\bf k}n'{\bf k}'}\}$ via the least squares fitting by $\tilde{W}_{n{\bf k}n'{\bf k}'}({\rm i}\omega)$. The approximate function $\tilde{W}_{n{\bf k}n'{\bf k}'}({\rm i}\omega)$ derived in this procedure, as well as the analytically continued ${\rm Re}\tilde{W}_{n{\bf k}n'{\bf k}'}(\omega\!+\!{\rm i}0^{+})$, well reproduces the calculated $W_{n{\bf k}n'{\bf k}'}({\rm i}\omega)$ and ${\rm Re}W_{n{\bf k}n'{\bf k}'}(\omega\!+\!{\rm i}0^{+})$, respectively (see Fig.~\ref{fig:fit}). For numerical stability, we have imposed a constraint $a_{i;n{\bf k}n'{\bf k}'}$$\geq$$0$ (Ref.~\onlinecite{comment-constraint}) and fixed $a_{i;n{\bf k}n'{\bf k}'}$ to $0$ for the plasmon modes with peaks much lower than the highest peak for each $n{\bf k}n'{\bf k}'$. We have considered up to two plasmon modes by setting $N_{\rm p}$$=$$2$~~\cite{Hoo-Hopfield,Sturm-Oliveira1989,Karlsson-Aryasetiawan, Silkin2007}.
\begin{table}[b!]
\caption[t]{Our calculated $T_{\rm c}$ considering only the phonon contributions (ph) to the exchange-correlation kernels, the phonon and static electron contributions (stat), and all the contributions ($N_{\rm p}$=2 and $N_{\rm p}$=1). Parameters $\lambda$$=$$2\int d\omega \alpha^{2}F(\omega)/\omega$ and $\omega_{\rm ln}$$=$${\rm exp}[\frac{2}{\lambda}\int d\omega {\rm ln}\omega\alpha^{2}F(\omega)/\omega]$ were calculated from the Eliashberg functions $\alpha^{2}\!F$~\cite{Migdal-Eliashberg}, and $r_{s}$$=$$\sqrt[3]{3/(4\pi \rho)}$ and $\Omega_{\rm p}$$=$$\sqrt{4\pi\rho/m^{\ast}}$ were calculated using the electron density $\rho$ and the band effective mass $m^{\ast}$, with $m^{\ast}$ evaluated from the fitting of the calculated DOS by that of the parabolic band.}
\begin{center}
\label{tab:Tc}
\begin{tabular}{lccccc} \toprule[2pt]
 & Al  &\multicolumn{4}{c}{fcc Li} \\
 &  &14GPa & 20GPa & 25GPa& 30GPa \\
$\lambda$ &0.417 &0.522 &0.623 &0.722 &0.812 \\
$\omega_{\rm ln}$ [K] & 314 &317&316&308&304 \\
$r_{s}$ &2.03 &2.71 &2.64  &2.59 &2.55 \\
$\Omega_{\rm p}$[eV] &16.2& 8.23& 8.44 &8.51 &8.58 \\
\midrule[0.5pt]
$T_{\rm c}^{\rm ph}$ [K]    &5.9 &10.0 &15.2 &19.0 &23.3  \\
$T_{\rm c}^{\rm stat}$ [K]  &0.8 &0.7  &1.8  & 3.2 &5.0           \\
$T_{\rm c}^{N_{\rm p}=2}$ [K]  &1.4 &2.2  &4.4 &6.8  &9.1 \\
$T_{\rm c}^{N_{\rm p}=1}$ [K]  &1.4 &2.2  &4.1 &6.5  &9.1          \\
$T_{\rm c}^{\rm expt.}$ [K]  & 1.20$^{a}$ & $<$4 &\multicolumn{3}{c}{5        --        17} \\
\midrule[2pt]
\multicolumn{6}{l}{$^{a}$Ref.~\onlinecite{Ashcroft-Mermin}}\\
\end{tabular} 
\end{center}
\end{table}

{\it -Application.} We performed the calculations for fcc lithium under pressures of 14, 20, 25, and 30GPa. All our calculations were performed within the local-density approximation~\cite{Ceperley-Alder,PZ81} using {\it ab initio} plane-wave pseudopotential calculation codes {\sc Quantum Espresso}~\cite{Espresso,Troullier-Martins} (see Ref.~\onlinecite{comment-detail} for the detail). The lattice parameter used for each pressure was derived from the Murnaghan equation of state~\cite{Murnaghan} with total energy calculations for different lattice parameters. Phonon frequency and electron-phonon coupling were calculated by the density functional perturbation theory~\cite{Baroni-review}. The dielectric functions considering the electronic screening were calculated within the RPA. The phonon contributions of the SCDFT exchange-correlation kernels ($\mathcal{K}^{\rm ph}$ and $\mathcal{Z}^{\rm ph}$) were calculated using the energy-averaged approximation~\cite{GrossII}, whereas the electron contributions ($\mathcal{K}^{\rm el,stat}$ and $\Delta\mathcal{K}^{\rm el}$) were calculated by Eq.~(13) in Ref.~\onlinecite{Massidda} and Eq.~(\ref{eq:Delta-kernel}). The SCDFT gap equation was solved with a random sampling scheme given in Ref.~\onlinecite{Akashi-MNCl}, with which the sampling error in the calculated $T_{\rm c}$ was not more than a few percent. We also performed the calculation for aluminum in ambient pressure for comparison.

\begin{figure}[htbp]
 \begin{center}
  \includegraphics[scale=.65]{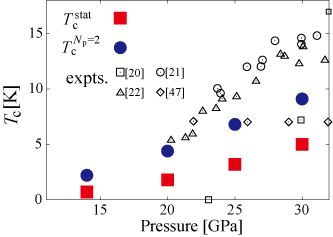}
  \caption{(Color online) Our calculated $T_{\rm c}$ (solid squares and circles) for fcc lithium under high pressures compared with the experimentally observed values. The open symbols represent the experiments: Ref.~\onlinecite{Shimizu2002} (open squares), Ref.~\onlinecite{Struzhkin2002} (open circles), Ref.~\onlinecite{Deemyad2003} (open triangles), and Ref.~\onlinecite{Lin-Dunn} (open diamonds).}
  \label{fig:Tc-expt}
 \end{center}
\end{figure}

Table \ref{tab:Tc} lists our calculated $T_{\rm c}$ with $\mathcal{K}$$=$$\mathcal{K}^{\rm ph}$ ($T_{\rm c}^{\rm ph}$), $\mathcal{K}$$=$$\mathcal{K}^{\rm ph}$$+$$\mathcal{K}^{\rm el,stat}$ ($T_{\rm c}^{\rm stat}$), and $\mathcal{K}$$=$$\mathcal{K}^{\rm ph}$$+$$\mathcal{K}^{\rm el,stat}$$+$$\Delta\mathcal{K}^{\rm el}$ ($T_{\rm c}^{N_{\rm p}=2}$). For examining the numerical stability of the fitting, the $T_{\rm c}$ values calculated with $N_{\rm p}$$=$$1$ ($T_{\rm c}^{N_{\rm p}=1}$) are also shown and they are almost the same as the $T_{\rm c}^{N_{\rm p}=2}$ values. We also show the electron-phonon coupling coefficient $\lambda$, the logarithmic average of phonon frequencies $\omega_{\rm ln}$, the density parameter $r_{s}$, and typical plasma frequency $\Omega_{\rm p}$. By using broad smearing functions~\cite{comment-detail}, we obtained $\lambda$ very similar to that of the latest calculation~\cite{Bazirov-pressure}, which is smaller than the earlier estimates~\cite{Tse,Kusakabe2005, Profeta-pressure,Kasinathan}. The material and pressure dependence of the experimentally observed $T_{\rm c}$ is already reproduced in the calculation with $\mathcal{K}$$=$$\mathcal{K}^{\rm ph}$, which can be understood in terms of the dependence of $\lambda$. By including $\mathcal{K}^{\rm el,stat}$, $T_{\rm c}$ significantly decreases, and $T_{\rm c}$ increases again when $\Delta\mathcal{K}^{\rm el}$ is introduced. The origin of the systematic increase due to $\Delta\mathcal{K}^{\rm el}$ is discussed in the following. The marked difference of the dynamical contribution between Al and Li is simply explained by the difference of $\Omega_{\rm p}$: The characteristic structure in $\Delta\mathcal{K}^{\rm el}$ (given later) becomes relevant as $\Omega_{\rm p}$ is lowered toward the phonon energy scale.

\begin{figure}[htbp]
 \begin{center}
  \includegraphics[scale=.6]{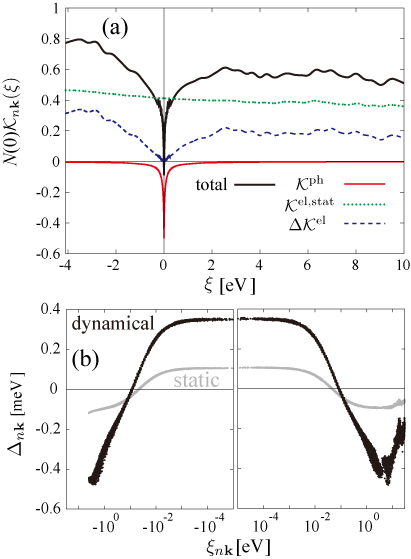}
  \caption{(Color online) (a) Contributions to the nondiagonal exchange-correlation kernel $\mathcal{K}_{n{\bf k}n'{\bf k}'}$ at $T$$=$$0.01$K calculated for fcc lithium under pressure of 14GPa, averaged by equal-energy surfaces for $n'{\bf k}'$. (b) The corresponding gap function calculated with (darker) and without (lighter) $\Delta\mathcal{K}^{\rm el}$.}
  \label{fig:kernel-gap}
 \end{center}
\end{figure}

We show in Fig.~\ref{fig:Tc-expt} the calculated $T_{c}$ for fcc Li together with the experimental values. By including $\Delta\mathcal{K}^{\rm el}$, the $T_{\rm c}$ increases up to the range of experimental values (from solid square to solid circle). In contrast, in Al the agreement with the experiment is well achieved with $\mathcal{K}$$=$$\mathcal{K}^{\rm ph}$$+$$\mathcal{K}^{\rm el,stat}$, and $\Delta\mathcal{K}^{\rm el}$ is irrelevant (see Table I). These results indicate the following: First, the high $T_{\rm c}$ in fcc Li under pressure is due to the help of the dynamical screening induced by plasmon, and second, our scheme works successfully regardless of whether their dynamical effects are strong or weak. 

In order to elucidate the origin of the increase of $T_{\rm c}$ by the dynamical effect, we examined energy-averaged nondiagonal kernels $\mathcal{K}_{n{\bf k}}(\xi)$$\equiv$$\frac{1}{N(\xi)}\sum_{n'{\bf k}'}\mathcal{K}_{n{\bf k}n'{\bf k}'}\delta(\xi-\xi_{n'{\bf k}'})$, where $N(\xi)$ is the electron density of states. With $n{\bf k}$ chosen as a certain point near the Fermi energy, we plotted the averaged kernel with $N_{\rm p}$$=$$2$ in Fig.~\ref{fig:kernel-gap} (a). We also decomposed the total kernel into $\mathcal{K}^{\rm ph}$, $\mathcal{K}^{\rm el,stat}$, and $\Delta\mathcal{K}^{\rm el}$. Within the phonon energy scale, the total kernel becomes slightly negative due to $\mathcal{K}^{\rm ph}$, whereas it becomes positive out of the phonon energy scale mainly because of $\mathcal{K}^{\rm el,stat}$. The $\Delta\mathcal{K}^{\rm el}$ value is positive definite, but nearly zero for a low energy scale. As a result, a valley-like structure in the total kernel around the Fermi energy is enhanced. Thanks to the retardation effect~\cite{Morel-Anderson}, this enhancement yields suppression of the effective repulsion between quasiparticles. Remarkably, the low-energy valley-like structure in $\Delta\mathcal{K}^{\rm el}$ appears in the energy range far smaller than the typical plasmon frequency (see Table \ref{tab:Tc}), which is consistent with the case of 3D electron gas reported by Takada~\cite{Takada1978}. 

The enhanced retardation effect depicted above can be seen more clearly from the gap functions plotted in Fig.~\ref{fig:kernel-gap} (b). In the SCDFT, the retardation effect is represented by sign inversion of the gap function in the energy region where $\mathcal{K}_{n{\bf k}}(\xi)$ becomes positive~\cite{GrossII}. The large negative gap value in the high energy region (darker points, $\gtrsim$1eV) shows a significance of the retardation effect from this region.

While it is expected that the dominant contribution of the dynamical screening is precisely evaluated by the RPA~\cite{Karlsson-Aryasetiawan,Silkin2007, Lazicki-Na}, further accuracy may be achieved by employing an improved exchange-correlation functional such as the hybrid functional~\cite{HSE-elemental}. To establish the plasmon effect in the present system, more detailed analysis of the gap function~\cite{Tse} is also important in future studies~\cite{comment-gap}.

{\it -Summary and conclusion.} We extended the SCDFT to the dynamical-RPA level so that the plasmons in solids are considered. Stable $T_{\rm c}$ calculations based on the present scheme were performed for fcc Li, with which the plasmon effect on $T_{\rm c}$ was quantified. We found that this effect substantially raises $T_{\rm c}$ by {\it cooperating} with phonon. In addition, we showed that the calculated $T_{\rm c}$ agrees better with the experimental values than that within the static-RPA level. Here, the plasmon-induced dynamical component of the screened Coulomb interaction enhances the retardation effect. Our present scheme provides a firm foundation for density functional theory for unconventional superconductivity induced/assisted by the electron correlation.

{\it -Acknowledgment.} The authors thank Kazuma Nakamura and Yoshiro Nohara for providing subroutines for calculating the RPA dielectric functions. This work was supported by Funding Program for World-Leading Innovative R\&D on Science and Technology (FIRST program) on ``Quantum Science on Strong Correlation", JST-PRESTO, Grants-in-Aid for Scientic Research (23340095) and the Next Generation Super Computing Project and Nanoscience Program from MEXT, Japan.


\begin{thebibliography}{999}
\bibitem{comment-constraint} This constraint is justified because the loss function is positive definite.
\bibitem{comment-gap} It should be noted that the gap function in SCDFT is an auxiliary quantity to calculate $T_{\rm c}$, so that one must be careful in comparing it directly with the experimentally observed superconducting gap.
\end{thebibliography}
\end{document}